\PassOptionsToPackage{colorlinks}{hyperref}
\PassOptionsToPackage{table, dvipsnames}{xcolor}
\documentclass[sigconf,screen]{acmart}

\usepackage{listings}
\usepackage{pifont}
\usepackage{graphicx}
\usepackage{algorithm}
\usepackage{algpseudocode}
\usepackage{booktabs, multirow} 
\usepackage{soul}
\usepackage{calc}
\usepackage{geometry}
\usepackage{enumitem}

\AtBeginDocument{\hypersetup{
    urlcolor=cyan}
    }

\newcommand{\pyt}{PyTorch}

\newcommand{\ld}{\texttt{Loader}}
\newcommand{\lds}{\ld\texttt{s}}
\newcommand{\inm}{\texttt{in-memory}}
\newcommand{\indexed}{\texttt{indexed}}
\newcommand{\irds}{\texttt{ir\_datasets}}
\newcommand{\dom}{{\color{OliveGreen}\ding{51}}}
\newcommand{\dontm}{{\color{red}\ding{55}}}

\definecolor{codegreen}{rgb}{0,0.6,0}
\definecolor{codegray}{rgb}{0.5,0.5,0.5}
\definecolor{codepurple}{rgb}{0.58,0,0.82}
\definecolor{backcolour}{rgb}{0.95,0.95,0.92}
\newlength{\Width}%
\newlength{\HeightReference}
\newlength{\DepthReference}
\newcommand{\MyColorBox}[2][red]%
{%
    \settowidth{\Width}{#2}%
    \colorbox{#1}%
    {%
        \raisebox{-\DepthReference}%
        {%
                \parbox[b][\HeightReference+\DepthReference][c]{\Width}{\centering#2}%
        }%
    }%
}
\lstdefinestyle{mystyle}{
  backgroundcolor=\color{backcolour},   
  commentstyle=\color{codegreen},
  keywordstyle=\color{magenta},
  stringstyle=\color{codepurple},
  basicstyle=\linespread{1.1}\ttfamily\footnotesize,
  captionpos=b,
  tabsize=2
}
\newcommand{\code}[1]{{\color{codegray}\small\texttt{#1}}}
\newcommand{\dpl}{\code{DataParallel}}
\newcommand{\dpls}{\code{DP}}
\newcommand{\ddp}{\code{DistributedDataParallel}}
\newcommand{\ddps}{\code{DDP}}

\AtBeginDocument{%
  \providecommand\BibTeX{{%
    \normalfont B\kern-0.5em{\scshape i\kern-0.25em b}\kern-0.8em\TeX}}}

\begin{document}
\fancyhead{}
    \thanks{This research has been supported by \textit{NWO} projects \textit{SearchX} (639.022.722) and \textit{Aspasia} (015.013.027).}

\copyrightyear{2022} 
\acmYear{2022} 
\acmConference[RENEUIR@SIGIR'22]{Workshop on Reaching Efficiency in Neural Information Retrieval @ SIGIR 202}{July 15, 2022}{Madrid, Spain}
\acmDOI{10.1145/3406522.3446012}
\acmISBN{978-1-4503-8055-3/21/03}

\title{Moving Stuff Around}
\subtitle{A study on the efficiency of moving documents into memory for Neural IR models}

\author{Arthur Câmara}
\email{A.BarbosaCamara@tudelft.nl}
\affiliation{%
  \institution{Delft University of Technology}  
  \city{Delft}  
  \country{Netherlands}  
}
\author{Claudia Hauff}
\email{c.hauff@tudelft.nl}
\affiliation{%
  \institution{Delft University of Technology}  
  \city{Delft}  
  \country{Netherlands}  
}
\begin{abstract}
When training neural rankers with Large Language models, a practitioner generally leverages multiple GPUs, which increases the amount of available VRAM, allowing larger batches, and decreasing training time. 
However, managing how data moves between disk (HDD), memory (RAM) and video memory (VRAM) is generally overlooked by most research implementations when training such data-hungry models.
Instead, most implementations use a naive approach of loading all documents from disk into memory, which delegates to the framework (e.g., PyTorch) the task of moving data into VRAM.
With the increasing sizes of datasets available, a natural question arises: \emph{Is this the optimal solution for optimizing training time?}
Therefore, we study the impact on performance (samples per second) and memory footprint (MBs allocated) of three approaches to moving documents in this hierarchy and how they scale with multiple GPUs. Specifically, we look into: (i) Pre-loading into memory, (ii) Reading from disk files with a lookup table and (iii) Reading from compressed files on disk using a memory map.
We show that when using some of the most common libraries for IR research, loading all documents into memory is not always the fastest option and also not feasible for more than a couple of GPUs, given its large memory footprint. Meanwhile, a good implementation of data handling is considerably faster and more scalable. Additionally, we demonstrate that using multi-processing instead of multi-threading can prevent problems caused by Python's Global Interpreter Lock (GIL).
Finally, we also discuss how popular techniques for improving loading times, like memory pining, multiple workers, and RAMDISK usage, can reduce the training time further with minor memory overhead.
\end{abstract}

\maketitle

\section{Introduction}
Current advances in Neural IR, mainly fueled by pre-trained, Large Language Models like BERT, have finally made deep learning a practical, and even expected, part of most current Information Retrieval research. However, even with pre-trained models for the most common corpora, like MSMarco, training (or rather, fine-tuning) a neural ranker is still commonplace, given a new dataset, new architecture, or even new training scheme. 

The problem of efficiently moving data from disk into memory is not exclusive to Information Retrieval. For example, a recent work by~\cite{Mohan2021Analyzing} shows that some deep learning models for vision and speech processing can waste between 10 and 70\% of their training time on I/O blocking operations.

\begin{figure}
\begin{center}
\includegraphics[width=0.9\columnwidth,keepaspectratio]{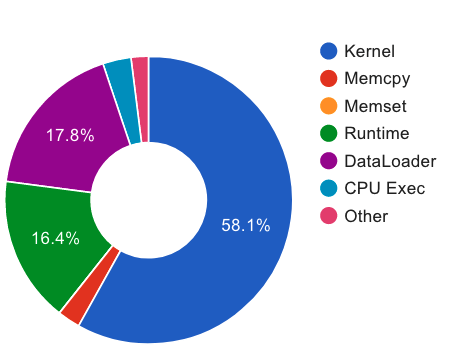}
\end{center}
  \caption{Tensorboard profiler for a CrossEncoder training run (500 steps) using the \texttt{ir-datasets} package for reading documents. $17.8\%$ of the runtime was spent on \texttt{DataLoader}, that is, reading and processing raw data from disk and sending to RAM}
  \label{fig:profiler}
\end{figure}

Much attention is devoted to creating the most parameter-efficient or faster training methods for neural models~\cite{Pradeep2022squeezing}. On the other hand, not much work has been dedicated to one of the most time-consuming steps when training such rankers: the time spent \emph{moving} data from disk into main memory. This step can take upwards of $15\%$ of the training time, as seen in Figure~\ref{fig:profiler}, and is mainly neglected, with most open-source implementations of neural rankers resorting to loading all the corpus documents into the main memory when starting the training process. 

As corpora grow more extensive, it quickly overcomes the available RAM, making it infeasible to load all documents into memory. Take, for instance, TREC's Disks 4 and 5~\cite{Voorhees1997TREC}, used by Robust04~\cite{Voorhees2004Robust}, a commonly used ranking dataset. In its uncompressed form, its size accounts for about 2GB. Meanwhile, MsMarco~\cite{nguyen2016msmarco}, the dataset used for TREC's Deep Learning Track in the last few years~\cite{Trec19, Craswell2020TREC20, Craswell21TREC}, takes over ten times that, at over 22GB, with its newer version, V2, surpassing the 100GB mark. Finally, other text corpora, like the Clean Common Crawl Corpus, used by the T5 model~\cite{Raffel2020Exploring}, can go even further, with more than 800GB needed to store the corpus.

The method of pre-loading all of the data from a corpus into memory, here called ~\inm{}, allows for simple document loading, with usually a single line of code being necessary for accessing a document in constant time, using a lookup table (\code{documents[doc\_id]}). 
However, as discussed here, this approach is not only impossible for large datasets but also \emph{slower} than reading from disk in most scenarios and does not scale with the number of GPUs. This is further exacerbated when following best programming practices (i.e. using one \textit{process} per GPU instead of one \textit{thread} per GPU).

Most available \pyt{} implementations of ranker models that use this method of pre-fetching documents also resort to using the \dpl{} (\dpls{}) module from \pyt{} when training with multiple GPUs. While convenient and easy to implement, this approach is not only \emph{not} recommended by \pyt{}'s official documentation\footnote{\url{https://pytorch.org/docs/1.11/notes/cuda.html\#cuda-nn-ddp-instead}}, it also spawns one thread for each GPU, which, given Python's nature and Global Interpreter Lock (\textit{GIL})~\footnote{\url{https://docs.python.org/3/glossary.html\#term-global-interpreter-lock}}, can cause racing conditions, which leads to considerate overhead when more than one thread tries to read from the same memory pool.

Instead, the preferred method to use when training with multiple GPUs is to use the built-in \pyt{} library for multi-processing (i.e. \ddp{} (\ddps{})~\cite{Li2020Pytorch}). Spawning one process for each GPU makes it possible to sidestep the GIL completely. Additionally, \ddp{} also allows a practitioner to divide the training between multiple nodes (i.e. multiple computers or hosts). It also allows the model parameters to be divided for larger models, making larger models that would not be trainable in a single host or GPU usable.

In this work, we argue that these two practices (using \inm{} for loading documents and \dpls{} for dividing the task among multiple GPUs), while easy to implement, \emph{should not be used} when training neural rankers with large amounts of data. Moreover, as we discuss further, these approaches lead to (i) code that is of lower quality (i.e. not attaining the best \pyt{} practices) and (ii) waste of resources, as methods that take less memory and have higher performance (i.e. can process more samples per second) are available.

We show that, with minor code changes and by being smarter when moving documents from disk, we can increase the performance of our training loops, greatly decrease memory usage, and increase the scalability of our models, making better use of the limited computing resources usually made available to academic researchers.

Therefore, we show how three different approaches to loading data (here referenced as \lds{}) differ in performance (i.e. samples processed per second) and memory usage (i.e. GB of main memory used) when working on 1, 2, 4 or 8 consumer-grade GPUs. More specifically, we compare (i) \texttt{in-memory} loading (i.e. pre-fetching all documents in memory), (ii) \texttt{indexed}, reading all documents directly from disk, while using a simple lookup table to locate document positions in files, and (iii) \texttt{ir-datasets}~\cite{macavaney:sigir2021-irds}, a toolkit for loading datasets specifically for Information Retrieval research, that uses highly efficiently ordered \code{numpy}~\cite{harris2020array} indexes in memory and compressed files in disk, that is shown to have a good performance when reading documents from disk. We also examine if common tricks for improving data loading latency, specifically \texttt{pin\_memory}, RAMDISK and increasing the number of threads dedicated to data loading speed up training in these scenarios. 

The code used in our experiments is available at ~\url{https://github.com/ArthurCamara/ir_efficiency}. Additionally, a WandB project dashboard with results is also publicly available at ~\url{http://wandb.ai/acamara/ir_efficiency}.

\section{Methodology}\label{sec:met}
This section outlines how we set up our experiments and the testing environment. We start with a brief description of the main libraries and classes used, followed by our machine configuration and how the example neural ranker was defined. We close this section by describing our implementations for \lds{} and how we organized our experiments regarding hyperparameters and different configurations. 

\subsection{Libraries}

While other deep learning frameworks exist, most open-source research implementations of neural rankers rely on the same handful of libraries. Therefore, we also use these popular Python libraries in our experiments, with their latest stable version. Namely, we use Python, PyTorch~\cite{pytorch}, and Hugging Face Transformers~\cite{Transformers}, all installed using \texttt{pip} (cf. \texttt{Pipfile} in Listing~\ref{pipfile} for the specific version numbers) sourcing from the default Python Package Index \texttt{PyPI}.

\hspace*{-\parindent}%
\begin{minipage}{\linewidth}
\begin{lstlisting}[language=Python, caption=Pipfile used for the experiments, label=pipfile]
[[source]]
url = "https://pypi.org/simple"
verify_ssl = true
name = "pypi"

[packages]
torch = "1.11.0"
transformers = "4.18"
accelerate = "0.6.2"
wandb = "0.12.15"
tqdm = "4.64.0"
psutil = "5.9.0"
ir-datasets = "0.5.1"

[requires]
python_version = "3.9"
\end{lstlisting}
\end{minipage}

When handling data, we kept as much as possible to standard \pyt{} classes. For example, our implementations for all dataset classes inherit from \code{Dataset}; we also use the standard \code{DataLoader} class for batching and splitting the data between GPUs~\footnote{Throughout this paper, we suppress the parents of the standard \pyt{} classes for simplicity}.

For our experiments with multiple GPUs, we use both the standard \dpl{} (\dpls{}) for \textit{multi-threaded} experiments and Hugging Face Accelerate\footnote{\url{https://github.com/huggingface/accelerate}} for \textit{multi-processed} experiments. It is generally preferred to run parallel programs in Python using multiple independent processes rather than one process with multiple threads. This is because Python, in its current state, is limited in its multi-threaded capabilities by its Global Interpreter Locker (GIL). 

The biggest drawback, and probably the main reason most research implementations don't adopt \ddps{}, is that doing so is not as simple as \dpls{}. 
It requires, in some cases, explicit synchronization and communication between different processes and, therefore, more changes to the training code when compared to \dpls{}. 
However, many libraries provide abstractions over \ddps{}, making it easier to use multi-processing without the added implementation complexities. Some of these libraries include the previously mentioned Accelerate, PyTorch Lightning~\footnote{\url{https://www.pytorchlightning.ai}} and Ray~\footnote{\url{https://www.ray.io}}.
It is worth mentioning that, according to the \pyt{} documentation\footnote{\url{https://pytorch.org/docs/1.11/generated/torch.nn.DataParallel.html}}, using \ddps{} is the preferred implementation when using multiple GPUs.

For tracking our experiments, we rely on Wandb~\cite{wandb}. It can track multiple hyperparameters and metrics in real-time while making it easy to evaluate multiple runs at the same time\footnote{The workspace with all runs and logging data can be accessed at \url{https://wandb.ai/acamara/ir_efficiency}}.

We ran all of our experiments in the same machine, equipped with a 56-core Intel Xeon E5-2690 v4 running at $2.60$GHz with $128$GB of DDR4 main memory running at 2400 MT/s. The datasets are stored on a 2TB SSD Drive, connected via a SATA $3.0$ connector. We also make use of 8 Nvidia GTX 1080 Ti. While slightly older and without Tensor Cores for accelerating Tensor-based operations, it is a still commonplace GPU for smaller laboratories and research groups. It also potentially makes the latency of loading data into memory less noticeable since the tensor operations may take longer. Therefore, the impacts reported here could be potentially more significant on more modern systems, with faster kernel operations but similar memory transfer speeds between disk and memory.

\subsection{\lds{} implementations}
We aim to cover a comprehensive set of possibilities when training a neural ranker model. However, it is not practical to cover every possible setup or scenario. Additionally, we are not interested in finding the best model effectiveness-wise but rather studying how the different options discussed here behave.
Therefore, we restrict our tests to one common ranker model and reuse it in all our experiments. We use a commonly used Cross-encoder~\cite{Yilmaz2019CrossDomainMO, MacAvaney2019CEDRCE,Nogueira2019PassageRW} with a DistilBERT-Base model~\cite{distil} that concatenates query and documents with a \texttt{[SEP]} token. The output \texttt{[CLS]} token embedding is fed to a 1-layer feed-forward network that estimates the relevance of the document w.r.t. the query. This model is also called BERT$_{CAT}$~\cite{TASB}.

We also train with a fixed batch size of $16$ per GPU, the largest size we could fit given our hardware. Each batch consists of triples of queries, positive and negative samples from the document ranking set from the TREC Deep Learning (DL) 2019 shared task~\cite{Trec19}, where the documents come from the MSMarco document dataset~\cite{nguyen2016msmarco}. The negative samples are randomly selected from all documents in the collection at runtime\footnote{Since we pre-load all documents identifiers in memory, the selection runs in constant time, and the document fetching stage performs according to each \ld{}.}.

We implement three different approaches for loading data into memory (\lds{}). Other possibilities---not discussed here---are using an in-memory database for storing documents, like Redis, or an inverse-index solution, like Anserini~\cite{Anserini} or Terrier~\cite{Pyterrier}. While the former could be helpful in a situation of high throughput or a production-ready scenario, we did not find a significant difference in keeping documents in memory in past experiments and therefore do not pursue it further. As for other IR toolkits, while highly optimized for high throughput and fast lookup, they can add significant overhead when looking for individual documents, as shown by~\citet{macavaney:sigir2021-irds}.

We now discuss our three \lds{} in turn.

\paragraph{\textbf{\inm{}}}
The most common approach in research code for neural rankers is to load all of the corpus into memory with a standard Python dictionary. This approach is easy to implement and use, generally requiring a single \code{for} loop over the documents for loading into memory and a single lookup in a dictionary that takes constant time.
However, this approach leads to high memory usage, especially when using multiple processes, or a considerable overhead when using multiple threads. When using \code{DistributedDataParallel} (i.e., multiple processes), each process loads a copy of the whole corpus on its own memory space during startup. When using multiple threads (i.e. \code{DataParallel}), the data remains in a shared memory space, which can lead to overhead and race conditions due to Python's GIL. In these cases, an alternative is to increase the number of workers in the \code{DataLoader} object, but that can trigger a copy of the whole dataset to each loader, which can increase memory usage linearly with the number of workers used.~\footnote{\href{https://github.com/UKPLab/sentence-transformers/blob/40af04ed70e16408f466faaa5243bee6f476b96e/examples/training/ms_marco/train_bi-encoder_margin-mse.py\#L91}{SBERT}~\cite{reimers-2019-sentence-bert},  \href{https://github.com/naver/splade/blob/eb74d1ab31c42b9b94df7c0d67cc582b2c972dda/anserini_evaluation/utils.py\#L35}{SPLADE~\cite{Formal2021_splade}} and  \href{https://github.com/sebastian-hofstaetter/matchmaker/blob/210b9da0c46ee6b672f59ffbf8603e0f75edb2b6/preprocessing/msmarco_makeqidpid.py\#L3}{MatchMaker}~\cite{Hofstaetter2020_sigir} are examples of open-source implementations that use~\dpls{}.}

\paragraph{\textbf{\indexed{}}}
As an alternative to \inm{}, \indexed{} loads only a pointer to each document byte-level address in the original file. Therefore, when requesting a document, this \ld{} looks into a small in-memory dictionary with the document's address in a file stored on disk.
If there already exists an opened \code{FileIO} object (i.e., an open file object), Python's \code{.seek()} function is $O(1)$ in run time, since it calls Linux's \href{https://man7.org/linux/man-pages/man2/lseek.2.html}{\code{lseek}} function, that executes in constant time. Therefore, reading a document into memory is only dependent on the size of the document, not the size of the corpus, as a naive lookup would be, with minimal memory overhead. A simplified version of this \ld{} is shown in Listing~\ref{alg:indexed}.

\paragraph{\textbf{\irds{}}} \texttt{ir\_datasets}~\cite{macavaney:sigir2021-irds} is a recent Python library that deals with a similar problem as we discuss here. The authors proposed a toolkit for abstracting common operations performed on datasets used for Information Retrieval research by simplifying downloading, storage and iteration over the documents, queries, and relevance labels for over 40 datasets. The dataset implements several optimizations for speed and quick experimentation, like lazily downloading datasets and a \code{docs\_store} API with a similar implementation as the above \indexed{}, but using \texttt{lz4} for compression of documents in disk and a \href{https://github.com/allenai/ir_datasets/blob/f21a9dedca6da205e4f6f63fa4676108af9145f6/ir_datasets/indices/numpy_sorted_index.py#L5}{sorted index offset}, built with memory maps from Numpy (\code{numpy.memap}) for quickly finding documents in disk. This is also similar to the implementation of Transformers-Rank\footnote{\url{https://github.com/Guzpenha/transformer_rankers}}, used in~\citet{Penha2021WeaklySL}. We expect this recently open-sourced library to be used more often in the future in combination with other toolkits. As a case in point, PyTerrier~\cite{Pyterrier} already \href{https://github.com/terrier-org/pyterrier/pull/93}{includes} \texttt{ir\_datasets} as part of the toolkit.

\subsection{Other optimizations}
Other possible optimizations that we also experiment with are mainly related to how \pyt{} deals with moving data from disk and main memory into the GPU's memory and are commonly accepted as good optimizations for \pyt{}\footnote{c.f. \pyt{}'s \href{https://pytorch.org/tutorials/recipes/recipes/tuning_guide.html}{documentation} on performance tuning and \href{https://nvlabs.github.io/eccv2020-mixed-precision-tutorial/}{this} NVidia's ECCV 2020~\cite{Vedaldi2020ComputerV} tutorial on Accelerating Computer Vision with Mixed Precision}.

First, we experiment with setting pinned memory (i.e., setting \code{pin\_memory=True} when defining a \code{DataLoader}). Pinning memory causes \pyt{} to pre-allocate a region of non-paged main memory for it, avoiding one implicit data copy inside the machine's main memory. Pinning memory also frees the CPU to fetch more data when done instead of waiting for the GPU operations to finish. 

Another commonly used option is for the \code{DataLoader} to use in multiple threads by setting the \code{num\_workers} parameter from the \code{DataLoader} object to a value greater than $0$ (its default value). Essentially, it allows the CPU to fetch data in multiple threads simultaneously without waiting for the GPU to finish its computations. 

Finally, we explore if we can mimic the fast access provided by the \inm{} approach by caching the files into RAMDISK. Using RAMDISK is a way to store data in the main memory while accessing it as a regular file. In practical terms, it means moving the corpus files from disk into a region in memory, usually on \texttt{/dev/shm} in most Linux distributions. In theory, moving the files into memory can mimic the speed benefits of \inm{} without its higher memory usage, since each thread or process would \emph{not} need to copy or re-read, respectively, the whole corpus into memory, as the OS sees the data as a regular file in disk.

\hspace*{-\parindent}%
\begin{minipage}{\linewidth}
\centering
\begin{lstlisting}[language=Python, caption=Simplified implementation for \indexed{}, label=alg:indexed]
def load_index(corpus_path:str):
    """Read all docs positions and store in-memory"""
    file_handler = open(corpus_path, "b")
    line = file_handler.readline()
    index = {}
    current_position = file_handler.tell()
    while line!=b"":
        line = line.decode("utf-8")
        doc_id, doc_text = line.split("\t")
        index[doc_id] = position
        current_position = file_handler.tell()
        line = file_handler.readline()
    return file_handler, index

def get_doc(index:dict, file_handler:FileIO, doc_id:str):
    """Return a doc given its id, a file and index"""
    position = index[doc_id]
    file_handler.seek(position) #O(1), calls lseek()
    return file_handler.readline().decode("utf-8")
    
\end{lstlisting}
\end{minipage}

\section{Results}

\begin{table*}[t]
\centering
\resizebox{\textwidth}{!}{
\begin{tabular}{lrrrrrrrrr}\toprule
\ld{} &\multicolumn{2}{c}{1 GPU} &\multicolumn{2}{c}{2 GPUs} &\multicolumn{2}{c}{4 GPUs} &\multicolumn{2}{c}{8 GPUs} \\\cmidrule(lr){2-3} \cmidrule(lr){4-5} \cmidrule(lr){6-7} \cmidrule(lr){8-9}
&\dpls{} &\ddps{} &\dpls{} &\ddps{} &\dpls{} &\ddps{} &\dpls{} &\ddps{} \\\midrule

\inm{} &39.56 &39.65 &72.33 &73.73 &140.47 &\cellcolor{OrangeRed!70}OOM &263.54 &\cellcolor{OrangeRed!70}OOM \\

\indexed{} &\cellcolor{Aquamarine!70}\ul{\textbf{39.88 (0.80\%)}} &\cellcolor{Aquamarine!70}\textbf{39.71 (0.15\%)} &\cellcolor{Thistle!70}72.09 (-0.34\%) &\cellcolor{Aquamarine!70}74.98 (1.69\%) &\cellcolor{Thistle!70}139.92 (-0.39\%) &140.04 (-) &\cellcolor{Thistle!70}262.56 (-0.37\%) &262.62 (-) \\

\irds{} &\cellcolor{Aquamarine!70}\ul{\textbf{39.88 (0.80\%)}} &\cellcolor{Thistle!70}39.16 (-1.25\%) &\cellcolor{Aquamarine!70}\textbf{72.78 (0.61\%)} &\cellcolor{Aquamarine!70}\ul{\textbf{75.45 (2.33\%)}} &\cellcolor{Aquamarine!70}\textbf{140.66 (0.13\%)} &\ul{\textbf{141.31 (-)}} &\cellcolor{Aquamarine!70}\ul{\textbf{268.29 (1.80\%)}} &\textbf{264.08 (-)} \\

\bottomrule
\end{tabular}
}
\caption{Number of samples/s processed by each \ld{} on 1, 2, 4 and 8 GPUs, using either \dpl{}(\dpls{}) (i.e. Multiple Threads) or \ddp{} (\ddps{}) (i.e. Multiple Processes). Values in (parenthesis) indicate the difference, in percentage, between that value and \inm{}. Results in \MyColorBox[Aquamarine!70]{blue} indicates a better performance when compared to \inm{}. Results in \MyColorBox[Thistle!70]{pink} indicates a worst performance. Results in \textbf{bold} indicates the better performance on that column. Results \ul{underlined} indicates the best result for that number of GPUs. Results in \MyColorBox[OrangeRed!70]{red} indicates that that experiment failed to run due to a Out Of Memory Error}
\label{tab:samples}
\end{table*}

In this section, we describe the results of the experiments described in Section~\ref{sec:met}. Recall that we are not interested in which method would yield better overall retrieval effectiveness (as they all share the same model and hyperparameters, it is unlikely that any significant difference would arise). Hence, we run each experiment for 1K training batches (between 16K and 128K query-document pairs in total, depending on the number of GPUs used) and report the average number of samples processed per second, as well as the total amount of memory used by the process. Note that, when running in multiple processes (i.e. using \ddps{} instead of \dpl{}), the reported memory usage from Python's \code{psutil} library is only for the main process, rather than for the whole program. Therefore, for a fair comparison, we multiply the reported memory usage by the number of GPUs used in that experiment. Also, note that we are \emph{not} considering the time each model takes to load the dataset. This step is only performed once and, given the relatively small number of steps performed, would bias the results against~\inm{}, as it naturally takes considerably longer to load the corpus into memory.

\subsection{Samples per second}

We begin by looking into different \lds{} described in Section~\ref{sec:met}, and how they perform when used with either \dpls{} or \ddps{}, with 1, 2, 4 or 8 GPUs. The results for how many samples per second each \ld{} can process is presented in Table~\ref{tab:samples}.

Perhaps surprisingly, we observe that the performance advantage of running with \inm{} is minimal to negative, especially as we increase the number of GPUs when using \ddps{}. And, if using \irds{} on two or more GPUs, its performance is better than \inm{} in every scenario. 
We believe this drop in performance of \inm{}, especially when using \dpls{}, has two leading causes. First, modern operating systems (OS) are highly optimized for reading data from disk, especially if using fast SSDs. Therefore, OS-level features, like pre-fetching, can play an important role here. But, even more interesting, the way that \dpls{} is implemented and how it splits batches between the GPUs is probably even more critical. 
Given Python's GIL and the multi-threaded nature of \dpls{}, the \inm{} object resides in a shared memory pool between each thread (i.e. each GPU has one thread by default). Therefore, when two or more threads request data at the same time, the GIL is used to control access to that pool with a single semaphore lock, only allowing one thread at a time to access this shared pool, which delays the access of multiple threads to the dataset. The same issue appears with \indexed{}, since Using optimizations like increasing the number of workers on the \code{DataLoader} initialization can help, as we show in Section~\ref{sec:extra}.

Another interesting finding is that, in most cases, using \ddps{} is slightly \emph{faster} than using \dpls{}, despite the communication overhead between processes. Again, we can blame the GIL. The more threads try to access the same memory pool, the more significant the bottleneck. The only noteworthy exception is \irds{} on 8GPUs. The reason is hard to pin down, but, looking at the other differences between \ddps{} and \dpls{}, it seems like the communication overhead between 8 processes actually made a significant impact. 

The drop in performance of \indexed{} on \dpls{} is also worth noticing. As more GPUs are added, the GIL impact is essentially doubled. First, there is a need to read from the shared lookup dictionary, which is not thread-safe. The GIL is, however, invoked a \textit{second} time, as reading from a file is also \textbf{NOT} thread-safe in Python. There are some possibilities to improve performance in this scenario, like using a \code{Queue} ~\footnote{\url{https://docs.python.org/3.9/library/queue.html\#Queue.Queue}}, but that would require considerable re-working of the code.

Finally, we note that, given its multi-processing nature, \ddps{} combined with \inm{} is not a feasible option for more than a couple of GPUs (at least for this dataset). With each process replicating the complete dataset in its memory space, we quickly run into Out-Of-Memory errors, with 4 and 8 GPUs crashing the machine before even starting the training process.

\subsection{Memory footprint}

\begin{table*}
\centering
\resizebox{\textwidth}{!}{
\begin{tabular}{lrrrrrrrrr}\toprule
\ld{} &\multicolumn{2}{c}{1 GPU} &\multicolumn{2}{c}{2 GPUs} &\multicolumn{2}{c}{4 GPUs} &\multicolumn{2}{c}{8 GPUs} \\\cmidrule(lr){2-3} \cmidrule(lr){4-5} \cmidrule(lr){6-7} \cmidrule(lr){8-9}

&\dpls{} &\ddps{} &\dpls{} &\ddps{} &\dpls{} &\ddps{} &\dpls{} &\ddps{} \\\midrule

\inm{} &40.23 &40.28 &40.84 &80.51 &42.14 &\cellcolor{OrangeRed!70}OOM &44.89 &\cellcolor{OrangeRed!70}OOM \\

\indexed{} &4.21 (-89.52\%) &4.26 (-89.44\%) &5.67 (-86.12\%) &8.45 (-89.50\%) &8.57 (-79.67\%) &16.84 (-) &13.96 (-68.91\%) &34.14 (-) \\

\irds{} &\ul{\textbf{2.79 (-93.07\%)}} &\textbf{2.83 (-92.97\%)} &\ul{\textbf{3.23 (-92.09\%)}} &\textbf{5.54 (-93.12\%)} &\ul{\textbf{4.34 (-89.69\%)}} &\textbf{11.20 (-)} &\ul{\textbf{6.72 (-85.03\%)}} &\textbf{22.59 (-)} \\

\bottomrule
\end{tabular}
}
\caption{Memory footprint (in GBs) for each \ld{} on 1, 2, 4 and 8 GPUs, using either \dpl{} (\dpls{}) (i.e. Multiple Threads) or \ddp{} (\ddps{}) (i.e. Multiple Processes). Results in \textbf{bold} indicates the better performance (lower memory usage) on that column. Results \ul{underlined} indicates the best result for that number of GPUs. Results in \MyColorBox[OrangeRed!70]{red} indicates that that experiment failed to run due to a Out Of Memory Error}
\label{tab:memory}
\end{table*}

A large dataset kept in memory would, naturally, quickly exhaust a machine's memory if replicated multiple times. However, if the performance can be increased with a suitable memory usage increase, the trade-off can be worth it. Therefore, in Table~\ref{tab:memory} we report the memory footprint for each \ld{}, parallelism method and number of GPUs. From that, it is clear that, overall, \irds{} has a significantly smaller memory footprint than either \inm{} or \indexed{}, regardless of parallelism method or number of GPUs.

Unsurprising, \inm{} has, by far, the largest memory footprint, with about 40GB allocated for each running process. Given the collection size on disk (22GB uncompressed), this is a significant increase over the original size. As expected, when using \ddps{}, memory usage grows linearly with each new instance, regardless of \ld{}. This, combined with the large footprint of \inm{} makes it impractical to be used in scenarios with more than a couple of GPUs, especially on a single-node training scheme, where each process would use memory from the same machine instead of sharing resources among multiple nodes.

However, when dealing with \dpls{}, \inm{}'s footprint does not seem to grow as fast as \indexed{}, or \irds{}. This lower-than-expected footprint  can be explained by the simplicity of the \inm{} implementation. By using a single, simple data structure (a \code{dict()}, that is essentially a hash map), it can easily share this same structure with all threads (albeit not in a thread-safe manner, given the GIL). 

Meanwhile, the growth shown by \indexed{}'s implementation appears to be triggering some caching by the OS, growing the footprint significantly faster than other methods (however, it still stays almost 70\% smaller than the footprint from \inm{}, even with 8 GPUs). As for \irds{}, its efficient implementation of Numpy's memory mapping greatly diminishes how fast its memory footprint grows for each extra thread.

Both \indexed{} and \irds{}, however, still increases its memory usage significantly faster than \inm{} when using \dpls{}. Due to their complexity, we believe this happens conversely to the simplicity of \inm{}'s data structures. Given that extra layer of indexes and optimized data formats, it is possible that either the OS or \pyt{}'s internals are replicating part of the data structure or pre-fetching documents from disk\footnote{We cannot find evidence, in \pyt{}'s documentation or code, that it does so explicitly. Therefore, we are inclined to believe this is due to some preemptive optimization by the OS.}.

\subsection{Common optimization strategies}\label{sec:extra}

\begin{table}
\resizebox{.47\textwidth}{!}{
\centering
\begin{tabular}{lrrrr}\toprule
Optimization Strategy &Value &Samples/s &Memory (GB) \\\midrule
\multirow{2}{*}{\code{pin\_memory}} &No &\textbf{141.62} &10.27 \\
&Yes &140.57 &\textbf{10.21} \\
\cmidrule(rl){1-4}
\multirow{4}{*}{\code{num\_workers}} &1 &\textbf{146.96} &10.00 \\
&2 &145.29 &10.00 \\
&4 &141.66 &10.00 \\
&8 &132.37 &\textbf{9.99} \\
\cmidrule(rl){1-4}
\multirow{4}{*}{\code{pin\_memory + num\_workers}} &1 &147.16 &\textbf{10.07} \\
&2 &147.09 &10.20 \\
&4 &147.13 &\textbf{10.07} \\
&8 &\ul{\textbf{147.26}} &10.15 \\
\cmidrule(rl){1-4}
\multirow{2}{*}{RAMDISK} &No &\textbf{141.62} &10.27 \\
&Yes &140.81 &\textbf{10.23} \\
\cmidrule(rl){1-4}
\code{pin\_memory} + 8 workers + RAMDISK & &147.23 &10.05 \\
\bottomrule
\end{tabular}
}
\caption{Impact of some of the most traditional optimization techniques for \pyt{} using \irds{} and \ddps{} on 4 GPUs. Results in \textbf{bold} indicate best performance for that type of strategy. The \ul{underlined} result is the best result overall. RAMDISK also adds the size of the corpus files (22GB in this case) to the total memory usage, as the files are copied to main memory}
\vspace{-0.5cm}
\label{tab:extra}
\end{table}

Given the popularity of \pyt{}, several optimization strategies are often suggested to increase its performance when loading data from disk. Therefore, we also explore how these may or may not impact performance for training neural rankers. Table~\ref{tab:extra} shows the impact of some of these techniques. For the sake of comparison, for this analysis, we fix the loader and parallel approach to the ones with best performance in Tables~\ref{tab:samples} and~\ref{tab:memory} (i.e., \irds{} on \dpls{}) and 4 GPUs. 

From the results, it is clear that most of them seem to have impacts ranging from negative, with deteriorated final results, to significant improvements, increasing performance even more than changing \ld{} or parallelism mechanism. For instance, using \code{pin\_memory} alone shows an decrease of 1.05 samples/s. However, increasing the number of processes dedicated to data loading from 0 (no parallelism) to 1 adds 5.3 samples/s, while, paradoxically, changing to 8 workers decreases the performance by 9.26 samples/s compared with no worker parallelism. Again, the GIL is to blame. While having one thread dedicated to loading data is useful, as the main thread can be dedicated to other tasks, adding more workers increases the probability that more than multiple threads try to access the same memory space simultaneously, running into race conditions.

Combining optimization, however, does seem to give a slight boost to performance. Using both \code{pin\_memory} and one worker increases the performance by 5.5 samples/s, slightly more than using \code{pin\_memory} alone, and the impact of adding extra workers does not seem to be negative, as without the pinned memory.

It is interesting to note that using RAMDISK (i.e., using main memory as disk space) decreases performance. Still, if used with other optimizations, line \code{pin\_memory} and more workers, it doesn't have a meaningful impact.

Memory-wise, the difference is minor, at less than 300MBs, since \irds{} already have a considerably low memory footprint and can probably be ignored in most cases.

\section{Conclusion and recommendations}
In this work, we presented how different methods for loading document corpora from disk into main memory behave in multiple scenarios. The main findings from our experiments are: (i) A naive implementation of loading the corpus in memory (i.e. the \inm{} loader) does not lead to performance gains, especially with more than one GPU (ii) Python's Global Interpreter Lock (GIL) can quickly impact performance when using more than one GPU and (iii) Using \ddp{} to sidestep the GIL leads to better scalability, but combining it with \inm{} is not a viable option for more than a few GPUs.

With these results, we have a few recommendations, presented as \textbf{Dos} (\dom{}) and \textbf{Don'ts} (\dontm{}) for practitioners and researchers wishing to train a neural ranker using \pyt{}. As shown, when adopted in conjunction, these should increase training performance and scalability of training procedures:

\begin{itemize}[leftmargin=*]
    \item[\dontm{}] \textbf{Don't load all of the documents in-memory}. As shown, \inm{} does not lead to any performance improvement. On the contrary, it is \textbf{slower} than other approaches, greatly increases memory usage, and does not allow the training to scale well to multiple GPUs. 
    \item[\dontm{}] \textbf{Don't use \dpl{}}. Not only is \dpls{} not recommender by \pyt{}'s documentation, it can quickly run into bottlenecks with Python's GIL and is slower than \ddps{}.
    \item[\dom{}] \textbf{Do use more efficient methods for reading documents from disk} and be aware that they may also be impacted by the GIL. In particular, \irds{}~\cite{macavaney:sigir2021-irds} seems to strike a good balance between ease of use and performance, while providing support to a large number of datasets in multiple formats.
    \item[\dom{}] \textbf{Do use \ddp{}, preferably with a wrapper library}. It is the recommended method by \pyt{}, it performs better than \dpls{} and can handle multiple training nodes if needed. To counter the extra code needed, wrappers, like Accelerate, can be used.
    \item[\dom{}] \textbf{Do use \code{pin\_memory} and increase \code{num\_workers} to at least 1}. While they may not improve much individually, combining common optimizations, specially setting only one extra thread for loading data, therefore avoiding the GIL, and setting \code{pin\_memory} to True can result in a sensible increase in performance.
\end{itemize}

\subsection{Caveats}
Naturally, we could not cover every possible combination of parameters or libraries. Rather, we choose to focus on simple approaches that can better extract extra performance when training neural rankers. 

Other strategies that can be useful, but were not studied here include: Setting gradients to \code{None} instead of using \pyt{}'s \code{zero\_grad} after each backward pass, and using AMP and FP16. The former can bypass some of the overhead of explicitly calling \code{zero\_grad} while having the same effect of zeroing the gradients of the network.The latter are two techniques that decrease the precision of the weights in the model, from 32bit to 16bit, and, in the case of AMP, allow for mixing between both automatically. The GPUs used for this experiment do not have support for this, and therefore, we could not test it. However, it's expected that this should lead to a considerable performance increase, since the lower precision can drastically decrease computation time and decrease VRAM usage, allowing for larger batches.

Our work is limited to \pyt{} and Hugging Face Transformers. While the most popular frameworks, especially for research, we recognize that other options exist, like TensorFlow and TF-Ranking~\cite{Kumar2018TFR}, that are widely used, especially in industry. 

Finally our conclusions regarding differences between \dpls{} and \ddps{} rely heavily on how Python currently implements multiple threads, specially the GIL. This may, however, change in the future, as exploratory work to remove the GIL is underway~\footnote{\url{https://pyfound.blogspot.com/2022/05/the-2022-python-language-summit-python_11.html}}. We are excited to see where this could lead, and what impact it may have on training performance. In theory, the removal of the GIL could mean the best of both worlds, with no performance degradation of multiple threads and lower memory usage, since there would be no need of data replication across workers. We note, however that, regardless of that, GIL or no GIL, a strategy like~\inm{} still requires that the corpus fits in main memory.

\bibliographystyle{ACM-Reference-Format}
\bibliography{references}

\end{document}